\documentstyle[12pt,psfig]{article}

\begin{document}

\title{The Microwave Background
Bispectrum  \\
Paper II: A Probe of the Low Redshift Universe}
\author{David M. Goldberg and David N. Spergel \\
Princeton University Observatory, Princeton, NJ\ 08544}
\maketitle

\begin{abstract}
Gravitational fluctuations along the line-of-sight from the surface of
last scatter to the observer distort the microwave background in
several related ways: The fluctuations deflect the photon path
(gravitational lensing), the decay of the gravitational potential
produces additional fluctuations (ISW effect) and scattering off of
hot gas in clusters produce additional fluctuations
(Sunyaev-Zel'dovich effect). Even if the initial fluctuations
generated at the surface of last scatter were Gaussian, the
combination of these effects produce non-Gaussian features in the
microwave sky. We discuss the microwave bispectrum as a tool for
measuring a studying this signal.  For MAP, we estimate that these
measurements will enable us to determine the fraction of ionized gas
and to probe the time evolution of the gravitational potential.
\end{abstract}

\section{Introduction}

Over the past few years, cosmologists have emphasized that
measurements of the two-point correlation function of CMB anisotropies
can yield a wealth of information about physical conditions in the
early universe. In this article, we will discuss how non-linear
physics can produce a detectable non-Gaussian signature in the
microwave background.  We concentrate on cases in which gravitational
lensing between here and the surface of last scatter couple via a
Limber's equation to produce a non-Gaussian signal.  The amplitude of
this signal is sensitive to the evolution of gravitational potential
fluctuations at low redshift. Thus, measurements of the
non-Gaussianity will complement measurements of the two point
correlation function by providing information about physical
conditions in the low-redshift universe.

In \S 2, we discuss cross-correlating the effects of gravitational
lensing and the Integrated Sachs-Wolfe (ISW) effect and
cross-correlating the effects of gravitational lensing with the
Sunyaev-Zel'dovich (SZ) effects.  Because there are equal numbers of
hot and cold spots, gravitational lensing alone does not produce a
detectable three-point signal\cite{Bernardeau98}.  However, the
combination of lensing with other low redshift effects will produce an
observable non-Gaussian signature. In \S 3, we compute the
signal/noise for this effect and discuss these cosmological
applications.  A companion paper\cite{Spergel99} provides a more
general discussion of the bispectrum, including derivation of the
angle-averaged bispectrum, a method for measuring the bispectrum and
estimating the signal to noise, and a fully worked example of the
bispectrum produced by the Rees-Sciama effect.

\section{Computing the Bispectrum}

The angle-averaged bispectrum is the spherical harmonic transform of
the temperature three point function,
\begin{eqnarray}
B_{l_1 l_2 l_2}&=&\sum_{m_1 m_2 m_3}
\left(
\begin{array}{ccc}
l_1 & l_2 & l_3 \\ 
m_1 & m_2 & m_3
\end{array}
\right)
\int d\hat{\bf l} d\hat{\bf m} d\hat{\bf n}
\langle T(\hat{\bf l}) T(\hat{\bf m}) T(\hat{\bf n}) \rangle \nonumber \\
&\times&Y_{l_1 m_1}^\ast(\hat{\bf l})
Y_{l_2 m_2}^\ast(\hat{\bf m})Y_{l_3 m_3}^\ast(\hat{\bf n})\ ,
\label{threept}
\end{eqnarray}
where $T(\hat{\bf n})$ is defined as the temperature anisotropy (with mean
zero) at a particular point in the sky.  We expand the temperature out into
three terms:
\begin{equation}
T(\hat{\bf n})=\sum_{lm} a_{lm}Y_{lm}(\hat{\bf n})
+\sum_{lm}a_{lm}\nabla\Theta(\hat{\bf n})\cdot\nabla Y_{lm}(\hat{\bf n})
+T_{lr}(\hat{\bf n})
\label{threeterms}
\end{equation}
where the first term is the CMB\ fluctuations from the surface of last
scatter, the second term describes the effects of lensing and the
third term describes the effects of secondary processes at low
redshift.  We shall assume throughout this paper that the third term
is first order (e.g. $\propto \Phi_0({\bf k})$), but clearly, second
order secondary processes would also give rise to an observable
bispectrum, as is the case with the Rees-Sciama effect considered in
the companion paper\cite{Spergel99}.  For this article, we will
consider two sources of secondary anisotropy (both of which are first
order), the ISW effect and the Sunyaev- Zel'dovich effect.

Combining equations (\ref{threept}) and (\ref{threeterms}) yields, 
\begin{eqnarray}
B_{l_1 l_2 l_2}&=&\sum_{m_1 m_2 m_3}
\left(
\begin{array}{ccc}
l_1 & l_2 & l_3 \\ 
m_{1_{}} & m_2 & m_3
\end{array}
\right)
\int d\hat{\bf l} d\hat{\bf m} d\hat{\bf n}
Y_{l_1 m_1}^\ast(\hat{\bf l})
Y_{l_2 m_2}^\ast(\hat{\bf m})Y_{l_3 m_3}^\ast(\hat{\bf n})
\nonumber \\
&\times&\sum_{ll'mm'}
Y_{lm}(\hat{\bf l})
\langle a_{lm}a_{l'm'}^\ast\rangle
\langle \nabla\Theta(\hat{\bf m})\cdot\nabla Y_{l'm'}^\ast(\hat{\bf m})
T_{lr}(\hat{\bf n})
\rangle
\end{eqnarray}
plus an additional 5 terms representing the various permutations of the
observation angles.  

We can immediately integrate over $\hat{\bf l}$:
\begin{eqnarray}
B_{l_1l_2l_3} &=&\sum_{m_1m_2m_3}\left( 
\begin{array}{ccc}
l_1 & l_2 & l_3 \\ 
m_1 & m_2 & m_3
\end{array}
\right)
\int d\hat{\bf m} d\hat{\bf n} Y_{l_2 m_2}^\ast(\hat{\bf m})
Y_{l_3 m_3}^\ast(\hat{\bf n})\\ \nonumber
&\times& c_{l_1}
\langle \nabla\Theta(\hat{\bf m})\cdot\nabla Y_{l_1 m_1}^\ast(\hat{\bf m})
T_{lr}(\hat{\bf n})
\rangle
\end{eqnarray}
We may note that by integrating by parts:
\begin{eqnarray}
\int d\hat{\bf m}\nabla^2 Y_{l_3 m_3}^\ast(\hat{\bf m})
\Theta(\hat{\bf m}) Y_{l_1 m_1}^\ast(\hat{\bf m})
&=&-\int d\hat{\bf m}
\nabla Y_{l_2 m_2}^\ast(\hat{\bf m})\cdot \nabla \Theta(\hat{\bf m})
Y_{l_1 m_1}^\ast(\hat{\bf m})\nonumber \\
&-&\int d\hat{\bf m}
\nabla Y_{l_2 m_2}^\ast(\hat{\bf m})\cdot \nabla Y_{l_1 m_1}^\ast(\hat{\bf m})
\Theta(\hat{\bf m})
\end{eqnarray}
where the surface integrals drop out, since we are integrating over a complete
unit sphere.  Combining this expression, with similar expressions involving the
Laplacian of $\Theta(\hat{\bf m})$ and $Y_{l_1 m_1}^\ast(\hat{\bf m})$, we
find: 
\begin{eqnarray}
B_{l_1l_2l_3} &=&\frac 12\sum_{m_1m_2m_3}\left( 
\begin{array}{ccc}
l_1 & l_2 & l_3 \\ 
m_1 & m_2 & m_3
\end{array}
\right)
\int d\hat{\bf n}Y_{l_3 m_3}^\ast(\hat{\bf n})c_{l_1} \\ \nonumber
&\times&\int d\hat{\bf m} \left<
\left[
\nabla^2 Y_{l_2 m_2}^\ast(\hat{\bf m})\Theta(\hat{\bf m})
Y_{l_1 m_1}^\ast(\hat{\bf m}) \right. \right. \\ \nonumber
&-&\left. \left. Y_{l_2 m_2}^\ast(\hat{\bf m})\nabla^2 \Theta(\hat{\bf m})
Y_{l_1 m_1}^\ast(\hat{\bf m})
- Y_{l_2 m_2}^\ast(\hat{\bf m})\Theta(\hat{\bf m})
\nabla^2 Y_{l_1 m_1}^\ast(\hat{\bf m})
\right]T_{lr}(\hat{\bf n})
\right>
\label{bb}
\end{eqnarray}
Since $\Theta (\hat{\bf m)}$ and $T_{lr}(\hat{\bf n})$ are scalar quantities,
their expectation value will depend only on their spatial separation:
\begin{eqnarray}
\langle \Theta(\hat{\bf m})T_{lr}(\hat{\bf n})\rangle&=&
\sum_{l}
\left(\frac{2l+1}{4\pi}\right)b_lP_l(\hat{\bf m}\cdot\hat{\bf n}) \\ \nonumber
&=&\sum_{lm} b_l Y_{lm}^\ast(\hat{\bf m}) Y_{lm}(\hat{\bf n})
\label{corr}
\end{eqnarray}
Inserting equation (\ref{corr}) into equation (\ref{bb}) and integrating
yields, 
\begin{eqnarray}
B_{l_1l_2l_3}&=&\frac{l_2(l_2+1)-l_1(l_1+1)-l_3(l_3+1)}2\sqrt{\frac{%
(2l_1+1)(2l_2+1)(2l_3+1)}{4\pi }} \nonumber \\ &&\times \left( 
\begin{array}{ccc}
l_1 & l_2 & l_3 \\ 
0 & 0 & 0
\end{array}
\right) c_{l_1}b_{l_3}
\label{eq:bispec}
\end{eqnarray}
Plus 5 additional terms, reflecting the various permutations of $l_1$, $l_2$
and $l_3$.
\subsection{Cross-correlating Lensing with the ISW Effect}

If the universe is not flat with $\Omega_{m}=1$, then at late times,
gravitational potential fluctuations decay. This decay generates additional
microwave background fluctuations: 
\begin{eqnarray}
T^{ISW}(\hat {{\bf n}}) &=&
2\int_0^{\tau _0} d\tau \dot \phi (\tau ,\hat {{\bf n}}\tau )  \nonumber \\
&=&2\int_0^{\tau _0}d\tau \frac{d^3{\bf k}}{\left( 2\pi \right) ^3}
\Phi_0({\bf k})\exp (i{\bf k}\cdot \hat {{\bf n}}\tau )
\dot\phi(\tau )
\end{eqnarray}
where the gravitational potential fluctuations have been expanded out in
Fourier space, $\hat {{\bf n}}$ is the direction of photon propagation, and
$\tau$ is conformal (comoving) lookback time. If a line-of-sight passes
primarily through high density regions, then it will be on average hotter than
lines-of-sight that pass primarily through low density regions.

Foreground fluctuations in the gravitational potential also distort the
photon geodesics, 
\begin{eqnarray}
\vec{\theta}(\hat{{\bf m}}) &=&
2\int_0^{\tau _0} d\tau \frac{\tau_0-\tau}{\tau _0}
\nabla_{\perp} \phi (\tau ,\hat{{\bf m}}\tau )   \\
&=&
2i \int d\tau
\frac{d^3{\bf k}}{\left( 2\pi \right) ^3} k \Phi_0({\bf k})
(\hat{{\bf k}}-(\hat{{\bf k}}\cdot \hat{{\bf m}})\hat{{\bf m}})
\exp (i{\bf k}\cdot\hat{{\bf m}}\tau )\phi (\tau )
\frac{\tau_0-\tau }{\tau _0}
\nonumber \\
&\equiv& \nabla_{\perp} \Theta(\hat{{\bf m}})\nonumber  \end{eqnarray}
This effect is surprisingly large; a typical photon's trajectory has been
deflected roughly several arcminutes from its initial path.

By measuring the three-point function, we can cross-correlate these two
effects,
\begin{eqnarray}
\langle T^{ISW}(\hat {{\bf n}}) \Theta(\hat{{\bf m}})\rangle
&=&{8 \over \pi}\int k^2 dk P({\bf k}) \int_0^{\tau _0}d\tau  
\int_0^{\tau _0} d\tau' \phi(\tau ){\frac{\tau_0-\tau }{\tau \tau _0}}
\nonumber \\
&&\times \dot\phi(\tau' ) \sum_{l,m} j_l(k\tau) j_l(k\tau')
Y_{lm}^{*}(\hat{\bf m})Y_{lm}(\hat{\bf n})\ ,
\end{eqnarray}
where we have used the Rayleigh expansion of the Fourier mode into
spherical harmonics:
\begin{equation}
e^{i{\bf k}\cdot{\bf r}}=4\pi\sum_{lm}i^lj_l(kr)Y_{lm}^\ast(\hat{\bf r})
Y_{lm}(\hat{\bf k})\ .
\end{equation}

\begin{figure}[tbp]
\centerline{\psfig{figure=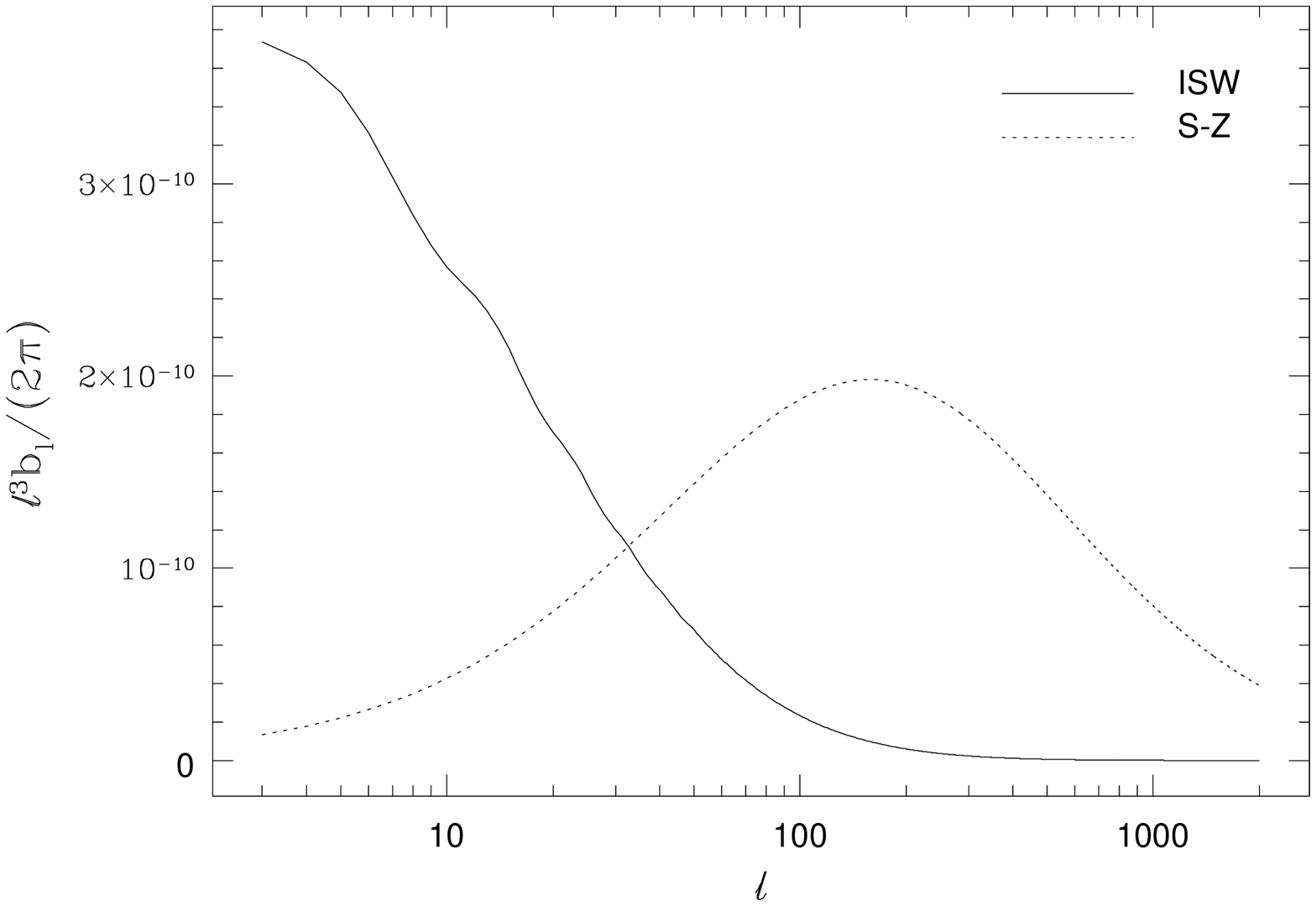,height=5.0in,angle=0}}
\caption{{}The $b_l$ coefficients as a function of $l$.  The solid line
shows the coefficient from the ISW effect, while the the dashed line
shows the coefficient arising from the Sunyaev Zel'dovich effect.
Each of the plots is for an $\Omega_m=0.3$, $\Omega_\Lambda=0.7$,
$h=0.65$ cosmology. We can estimate the contribution that the
different coefficients will have through dimensional analysis.  Since
$c_l\propto l^{-2}$, the Wigner 3-j symbols $\propto l^{-1/2}$
(equation~\ref{eq:bispec}) and the signal per $l_3$ is expected to go
as $B_{l_1 l_2 l_3}\times l^{-4}$ (see equation~\ref{eq:chi2}) the
$\chi^2/\Delta l_3$ will be roughly constant if $b_{l}\propto l^{-3}$.
Thus we have normalized the kernels with an $l^3$ prefactor in order
to estimate their importance.}
\label{fg:b_l}
\end{figure}

Since most of the contribution to the integral comes from when $\tau \simeq
\tau ^{\prime }$ , we approximate $\dot \phi $ by its value at that time and
then integrate over $\tau ^{\prime }$: 
\begin{equation}
b^{ISW}_l=-\frac 4{\pi ^{1/2}}
\frac{\Gamma \left( \frac {l}{2}+\frac{1}{2}\right)}
{\Gamma \left( \frac{l}{2}+1\right) }
\int P(k)k dk\int d\tau \phi (\tau)j_l(k\tau )\dot \phi(\tau)
\frac{\tau _0-\tau }{\tau \tau _0}
\label{aterm}
\end{equation}
In our analysis, we compute this integral numerically up to $l=50$.
However, for large $l$, we can evaluate this integral quickly by 
noting that  $j_l(k\tau ) \simeq \delta(k\tau - l)\sqrt{\pi/2l}$, yielding:
\begin{equation}
b^{ISW}_l \simeq -8 \int d\tau \phi(\tau) \dot{\phi}(\tau)
\frac{\tau_0-\tau}{\tau^{2}\tau_0} P(l/\tau)
\label{adelta}
\end{equation}
To illustrate the behavior of this coefficient, Figure~\ref{fg:b_l}
shows a plot of $b_l$ versus $l$ for an $\Omega_m=0.3$, $\Lambda=0.7$
cosmology.

\subsection{Cross-correlating Lensing with the Sunyaev-Zel'dovich Effect}

Filaments, groups, and clusters will distort the microwave background
through both the SZ effect and gravitational lensing\cite{Ostriker86}.
The amplitude of the effect depends upon the integrated pressure
fluctuations along the line-of-sight,
\begin{equation}
T^{SZ}(\hat {{\bf n}}) = -{2 \sigma_T\over m_e c^2} \int a d\tau \delta
p_e(\tau)
\end{equation}
where $\sigma_T$ is the Thompson cross-section, $m_e$ is the electron
mass and $\delta p_e$ is the fluctuations in electron pressure.

The electron pressure is the product of the electron density and the
electron temperature.  On large scales, we parameterize this as
$\delta p_e = b_{gas} \bar p \Delta$, where $b_{gas}$ is the gas bias
factor (which is expected to be $\sim 4$ as dense regions are hotter
than low density regions \cite{Refregier99}), $\bar p$ is the mean gas
pressure and $\Delta$ is the fluctuation in the dark matter density.
This approximation only holds on scales in which linear theory
holds. However, as we will show, the SZ-lensing bispectrum probes
scales of $k^{-1}>100\ {\rm Mpc}$.  Numerical simulations (R. Cen,
private communication) find that the density weighted temperature,
$kT_e \simeq 1/(1+z)$ keV for $z < 3$.  Since the gas density falls as
$(1+z)^{-3}$, the pressure drops roughly as $(1+z)^{-2}$.  Based on
this, we approximate the large scale SZ fluctuations as,
\begin{equation}
T^{SZ}(\hat {{\bf n}}) = -2 y_0 b_{gas} \int {d\tau \over \tau_0}
{\Delta(\hat {{\bf n}},\tau) \over a}
\label{SZ}
\end{equation}
where $y_0 = \sigma_T n_{e0} k T_{e0} \tau_0/m_e c^2$.  For a flat
universe with $H_0 = 65 km/s$, and $\Omega_{ionized} = 0.05$ $T_{e0} =
1$ keV, $y_0 = 1.2\times 10^{-5}$.  This approximation is likely to be
valid on scales larger than the non-linear scale.

We can relate the dark matter density to the potential and rewrite
equation (\ref{SZ}),
\begin{equation}
T^{SZ}(\hat{\bf n})=\frac{4 y_0 b_{gas}}{3H_0^2 \Omega_m}
\int \frac{d^3{\bf k}}{(2\pi)^3} k^2 \Phi_0({\bf k})
\int \frac{d\tau}{\tau_0}\phi(\tau)e^{i{\bf k}\cdot \hat{\bf n}\tau} 
\end{equation}
and now compute the $SZ-lensing$ cross-correlation term:
\begin{eqnarray}
\langle T^{SZ}(\hat {\bf n}) \Theta^\ast(\hat{\bf m})\rangle &=&
\frac{16y_0b_{gas}}{3\pi H_0^2\Omega_m}
\int dk k^4 P(k) 
\int d\tau d\tau' \phi(\tau) \phi(\tau') \frac{\tau_0-\tau}{\tau\tau_0^2}
\\ \nonumber
&\times&\sum_{lm}j_l(k\tau)j_l(k\tau') Y_{lm}^\ast(\hat{\bf m})
Y_{lm}(\hat{\bf n}) 
\end{eqnarray}
As $l$ gets large, the spherical Bessel functions can be treated as
approximating delta functions such that:
\begin{equation}
j_{l}(k\tau')\simeq \delta(l-k\tau')\frac{\sqrt{\pi}}{2}
\frac{\Gamma(\frac{l}{2}+\frac{1}{2})}{\Gamma(\frac{l}{2}+1)}
\end{equation}
Making this approximation for $\tau'$ and integrating from $-\infty$
to $\infty$ yields:
\begin{equation}
b_{l}^{SZ}\simeq
\frac{8 y_0 b_{gas}}{3\pi^{1/2}H_0^2\Omega_m}
\frac{\Gamma(\frac{l}{2}+{1}{2})}{\Gamma(\frac{l}{2}+1)}
\int dk \ k^3 P(k) \phi(l/k) \int d\tau \phi(\tau)
\frac{\tau_0-\tau}{\tau\tau_0^2}
\end{equation}

This may be further simplified by making the same approximation for
$j_l(k\tau)$ and integrating over $k$:
\begin{equation}
b_{l}^{SZ}\simeq
\frac{8 y_0 b_{gas}l^2}{3H_0^2 \Omega_m}
\int d\tau \phi^2(\tau)\frac{\tau_0-\tau}{\tau^5\tau_0^2}
P(l/\tau)
\label{SZeq}
\end{equation}
Note that for very small values of $\tau$ (low redshift), $P(l/\tau)\propto
\tau^{-7}$, and the kernel of the above equation goes to 0.  It is
only in the large wavelength regime in which $P(k)\propto k^{-5}$
where the integral is maximized.  This occurs around $\Lambda=100
h^{-1} {\rm Mpc}$ for most cosmological models, and thus, the
assumption that pressure and density perturbations are proportional
and evolve linearly are well motivated.

As Figure~\ref{fg:b_l} shows, the Sunyaev-Zel'dovich effect dominates
over the ISW effect at large $l$. Since the SZ effect is enhanced by
non-linear evolution, the estimated signal in equation (\ref{SZeq}) is
likely an underestimate, numerical simulations, complemented by
improved analytical techniques, will be needed to more accurately
compute the signal.

\begin{figure}[tbp]
\centerline{\psfig{figure=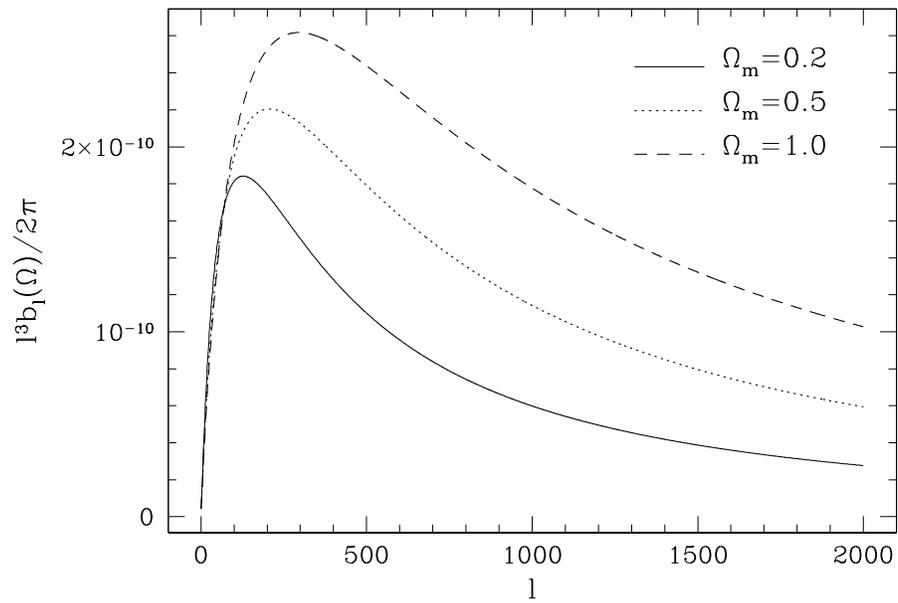,height=5.0in,angle=0}}
\caption{{}As in the previous figure, but the coefficients, $b_l$, are
shown only for the SZ-lensing coupling.  Here we have varied
$\Omega_m$ to illustrate that the shape of the coefficients, as well
as the normalization, vary with $\Omega_m$, and thus, $\Omega_m$ 
cannot be considered a degenerate parameter with the product, $y_0
b_{gas}$.}
\label{fg:b_l_omega}
\end{figure}

It may be thought that $b_l$ will merely change normalization with a
change of $\Omega_m$, and thus, $y_0 b_{gas}$ (which clearly only
affects the normalization) and $\Omega_m$ would be indistinguishable.
Since there is little variation in $c_l$ for variations of $\Omega_m$,
it might be thought that the bispectrum {\em shape} will be roughly
constant, and thus, only the normalization could be extracted.  In
Figure~\ref{fg:b_l_omega}, we show that this is not the case.  For
variations in $\Omega_m$, we see that $b_l$ varies not only in
normalization, but in shape as well, and thus, both $y_0 b_{gas}$ and
$\Omega_m$ will be able to be extracted almost independently.

\section{Discussion}

Will we be able to detect either effect? We can estimate this by determining
whether in a $\Omega _m=0.3,\Omega _\Lambda =0.7$ universe, we can use the
MAP data to reject the hypothesis that there is no non-Gaussianity in the
CMB maps: 
\begin{equation}
\chi^2=\sum_{l_1,l_2,l_3}{\frac{\langle B_{l_1,l_2,l_3}\rangle ^2}{\langle
B_{l_1,l_2,l_3}^2\rangle }}
\label{eq:chi2}
\end{equation}
where we restrict the sum to $l_1<l_2<l_3$.  For our standard set of
parameters, $\chi^2 \simeq 9$ with most of the signal coming from the SZ
effect.  Thus, measuring the bispectrum will enable us to measure
$\Omega_{ionized}$ and probe the time evolution of the gravitational
potential.

\begin{figure}[tbp]
\centerline{\psfig{figure=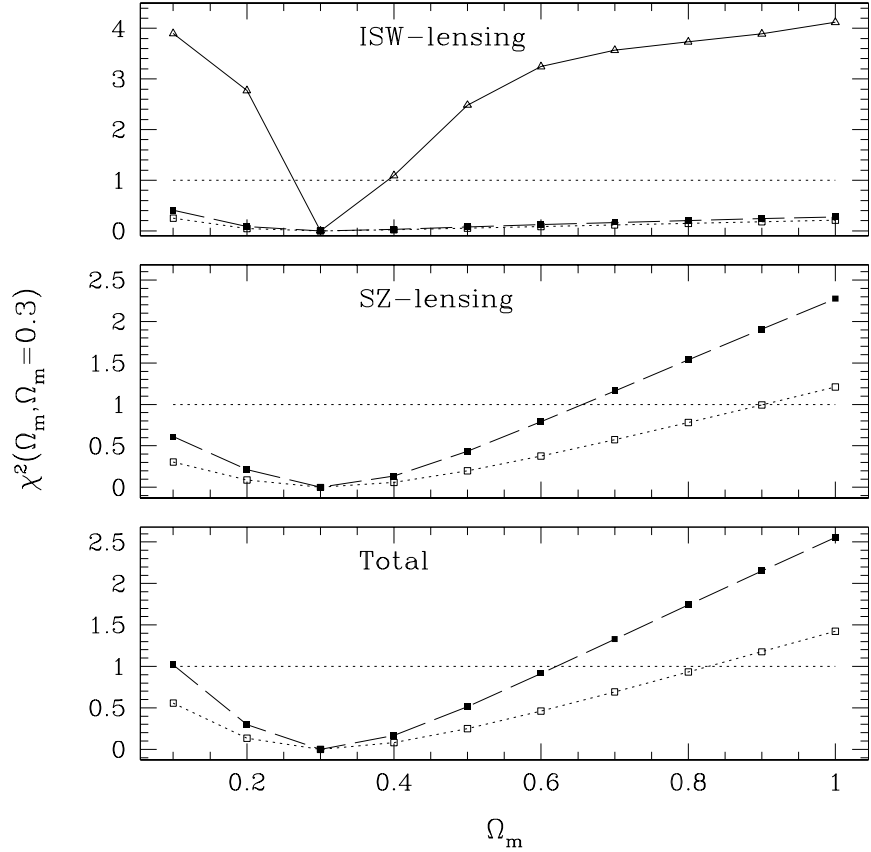,height=5.0in,angle=0}}
\caption{{}The $\chi^2$ differences in the ISW-lensing bispectrum, the
SZ-lensing bispectrum, and their sums between a fiducial model
$\Omega_m=0.3$, $\Omega_\Lambda=0.7$ and a test model with a flat
cosmology and varying $\Omega_m$.  The solid line shows the $\chi^2$
for the PLANCK experiment, while the dashed line shows $\chi^2$ for
MAP 4 year results and the the dotted line shows the MAP 1 year result.  
with 1 $\sigma$ uncertainty.}
\label{fg:chi2_omega}
\end{figure}

In Figure~\ref{fg:chi2_omega}, we plot the value of $\chi^2$ for the
ISW-lensing bispectrum for flat models with varying $\Omega_m$ and
$\Omega_\Lambda$.  As shown in the Figure, the PLANCK\cite{PLANCK}
satellite will do an excellent job differentiating between the null
hypothesis and a given value of $\Omega_m$, giving a 4 $\sigma$ signal
for $\Omega_m=1.0$.  The MAP satellite, however, will be unable to
differentiate between the two.

\begin{figure}[tbp]
\centerline{\psfig{figure=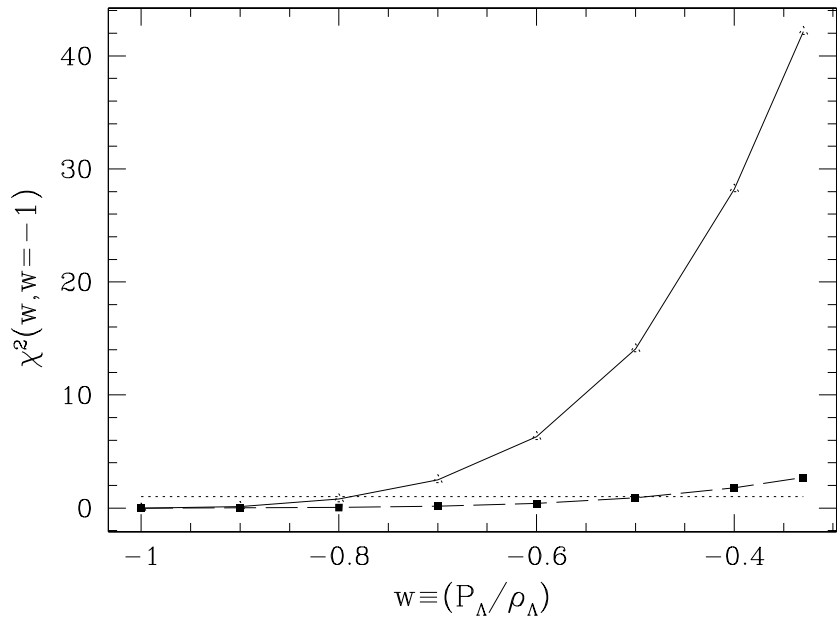,height=5.0in,angle=0}}
\caption{{}The same as the previous figure, but with variations in the
equation, $w$.  For a cosmological constant, $w=-1$ (our fiducial
model), while $w=-1/3$ is a curvature-like term.  Both MAP and PLANCK
will be able to distinguish between extreme cases; MAP would predict
$w<-0.4$ and PLANCK would give $w<-0.8$.}
\label{fg:chi2_isw_w}
\end{figure}

Likewise, this method will be able to potentially probe different
equations of state.  While the Cosmological constant behaves as a
component with  $w\equiv P/\rho=-1$ (our fiducial model),
different equations of state produce dramatically different rates of
potential growth, and hence, we may measure the $\chi^2$ difference
between test values of $w$ and $w=-1$, as shown in
Figure~\ref{fg:chi2_isw_w}.  Note that both MAP and PLANCK provide
constraints on $w$ at the 1 $\sigma$ level; PLANCK gives $w<-0.8$, and
MAP gives $w<-0.4$. 

As seen in Figure~\ref{fg:chi2_omega}, MAP will be sensitive to this
effect as well, and the 4 year results will yield 1 $\sigma$
uncertainties of $\delta\Omega_m=\pm 0.4$.  PLANCK can distinguish SZ
from CMB fluctuations, so it will be very sensitive to this effect as
well.  However, calculating its signal to noise is beyond the scope of
this paper.

\begin{figure}[tbp]
\centerline{\psfig{figure=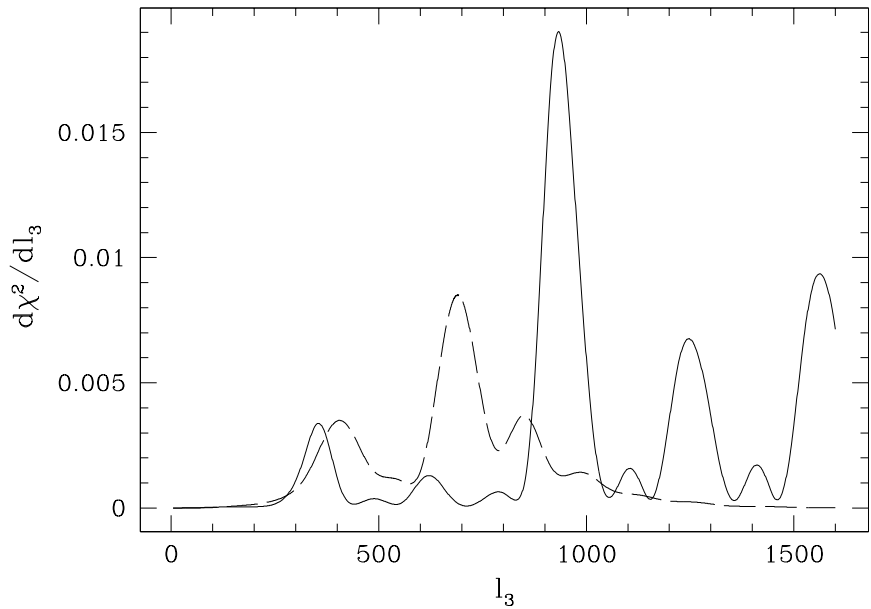,height=5.0in,angle=0}}
\caption{{}The $\chi^2$ per mode ($l_3$) as given by the bispectrum.
The solid line shows the results from the ISW-lensing effect measured
by PLANCK, while the dashed line shows the SZ-lensing effect as
measured by MAP (4 yr. results).  In general, the signal is at higher
wavenumber (smaller scale) for the SZ-lensing effect than for the
ISW-lensing effect.  However, PLANCK's increased sensitivity at higher
wavenumber distorts this relation in this plot.}
\label{fg:dchi2}
\end{figure}

Figure~\ref{fg:dchi2} shows the additional $\chi^2$ for a mode with a
given value of $l_3$ (by convention, the largest index).  Note that
the maximal contribution occurs near $l_3=900$ for PLANCK.  Though the
SZ-lensing effect is generally at higher mode numbers (see the
comparison between the signals as will be detectable by MAP), it will
nevertheless produce a much stronger signal in MAP than will the
ISW-lensing effect, since at high $l$ (but less than 600), the
SZ-lensing coefficient ($b_l$) dwarfs the coefficient for the
ISW-lensing effect (see Figure~\ref{fg:b_l}).

In this article, we estimated the non-Gaussian signature produced by
the combined effects of gravitational lensing and the ISW effect and
the combined effects of gravitational lensing and the SZ effect.  For
most cosmological models, the SZ signature is stronger and will swamp
the ISW effect signal.  Detecting the SZ-lensing cross-correlation
will determine the mean density of ionized gas
in the today's universe.  Thus, MAP should be able to detect the
``missing baryons"\cite{Fukugita98,Cen98}.

By cross-correlating the MAP data with observations of large scale
structure and X-ray maps, we can look for additional signatures of low
redshift physics.  Previous papers have discussed cross-correlating
the galaxy distribution in the Sloan Survey with the temperature maps
to look for gravitational lensing effects\cite{Suginohara98} and for
the Sunyaev-Zel'dovich produced by both clusters and
superclusters\cite{Refregier98}.  Boughn et al.\cite{Boughn98} have
looked for cross-correlations between the cosmic X-ray and microwave
backgrounds in an attempt to detect the late-time ISW effect.
Combining these different cross-correlations will enable cosmologists
to determine the relationship between the mass distribution and the
distribution of gas and galaxies.  These observations will also probe
the time evolution of the gravitational potential and provide a new
tool for measuring the basic cosmological parameters.

\section{Acknowledgments}

We thank Alex Refregier, Jeremy Goodman, Martin Bucher, Gary Hinshaw,
Arthur Kosowsky and Dick McCray for useful discussions. We also thank
Uros Seljak and Matias Zaldarriaga for sharing their results prior to
publication. DNS acknowledges the MAP/MIDEX project for support. DMG
is supported by an NSF graduate research fellowship.

\end{document}